\documentclass[%
reprint,
superscriptaddress,
a4paper,
showpacs,preprintnumbers,
amsmath,amssymb,
aps,
prstab,
floatfix,
]{revtex4-1}

\usepackage{graphicx}% Include figure files
\usepackage[caption=false]{subfig}
\usepackage{float}
\usepackage{hyperref}% add hypertext capabilities
\hypersetup{colorlinks=true,citecolor=blue}
\begin{document}
\preprint{APS/XX-XXX}
\title{Crystal slow extraction of positrons from DA$\Phi$NE: the SHERPA project}% Force line breaks with \\
\thanks{INFN CSN5 grant for young researchers}
\author{M.~Garattini}\email{PI, marco.garattini@lnf.infn.it}
\affiliation{INFN, Laboratori Nazionali di Frascati, C.P. 13, I-00044 Frascati, Italy}
\author{D.~Annucci}
\affiliation{ Dipartimento di Fisica, Sapienza Università di Roma,
P.le Aldo Moro 5, 00185 Roma, Italy}
\affiliation{INFN, Sezione di Roma, P.le Aldo Moro 2, I-00185 Roma, Italy}
\author{O.R.~Blanco-Garc\'ia}
\affiliation{INFN, Laboratori Nazionali di Frascati, C.P. 13, I-00044 Frascati, Italy}
\author{P.~Gianotti}
\affiliation{INFN, Laboratori Nazionali di Frascati, C.P. 13, I-00044 Frascati, Italy}
\author{S.~Guiducci}
\affiliation{INFN, Laboratori Nazionali di Frascati, C.P. 13, I-00044 Frascati, Italy}
\author{A.~Liedl}
\affiliation{INFN, Laboratori Nazionali di Frascati, C.P. 13, I-00044 Frascati, Italy}
\author{M.~Raggi}
\affiliation{ Dipartimento di Fisica, Sapienza Università di Roma,
P.le Aldo Moro 5, 00185 Roma, Italy}
\affiliation{INFN, Sezione di Roma, P.le Aldo Moro 2, I-00185 Roma, Italy}
\author{P.~Valente}
\affiliation{INFN, Sezione di Roma, P.le Aldo Moro 2, I-00185 Roma, Italy}
\date{\today}% It is always \today, today,
             %  but any date may be explicitly specified

\begin{abstract}
The SHERPA project aim is to develop an efficient technique to extract a positron beam from one of the accelerator rings composing the DA$\Phi$NE complex at the Frascati National Laboratory of INFN, setting up a new beam line able to deliver positron spills of O(ms) length, excellent beam energy spread and emittance. 
The most common approach to slowly extract from a ring is to increase betatron oscillations approaching a tune resonance in order to gradually eject particles from the circulating beam. 
SHERPA proposes a paradigm change using coherent processes in bent crystals to kick out positrons from the ring, a cheaper and less complex alternative. A description of this innovative non-resonant extraction technique is reported in this manuscript, including its performance preliminary  estimation.

\end{abstract}
\maketitle
%\tableofcontents
\section{Introduction}
%In the recent years there has been a growing interest for solutions to the dark matter problem assuming an almost completely secluded sector of new particles, with only very suppressed interactions with the ordinary matter. This kind of scenario opens the possibility of new feebly interacting particles in a wide range of masses, which can be produced in electron-positron annihilations. The availability of a high-intensity positron beam gives the bonus of a very high luminosity, at the price of a reduced available center-of-mass energy, in a fixed-target configuration. The PADME experiment~\cite{Raggi:2015gza} has been designed for searching this kind of light particles, in the form of a ``dark photon'' or a axion-like mediator, as a peak in the missing mass spectrum of mono-photons events in $e^+ e^- \to \gamma X$ annihilations exploiting the positron beam from LINAC of the DA$\Phi$NE complex. 
%
The PADME experiment~\cite{Raggi:2015gza} has been designed to search for a new kind of dark sector light particle, like a ``dark photon'' or an axion-like mediator, seen as a peak in the missing mass spectrum of mono-photon events in $e^+ e^- \to \gamma X$ annihilations of positrons on target. Very high luminosity is achieved in a fixed-target collision scheme, exploiting the positron beam coming from the LINAC of the DA$\Phi$NE complex~\cite{Boni:1998qe}, albeit at the price of a reduced center-of-mass energy below 20~MeV.\par
At DA$\Phi$NE~\cite{Vignola:1993dy}, one beam of electrons and one of positrons collide at the $\phi$ meson resonance (1.02~GeV), where each beam is composed by a maximum of 120 bunches stored in two Main Rings (MR). The beams are populated by a high-current LINAC and a small damping ring that reduces the emittance of the transported beam.\par
\begin{figure*}[htb]
  \includegraphics[width=0.96\textwidth]{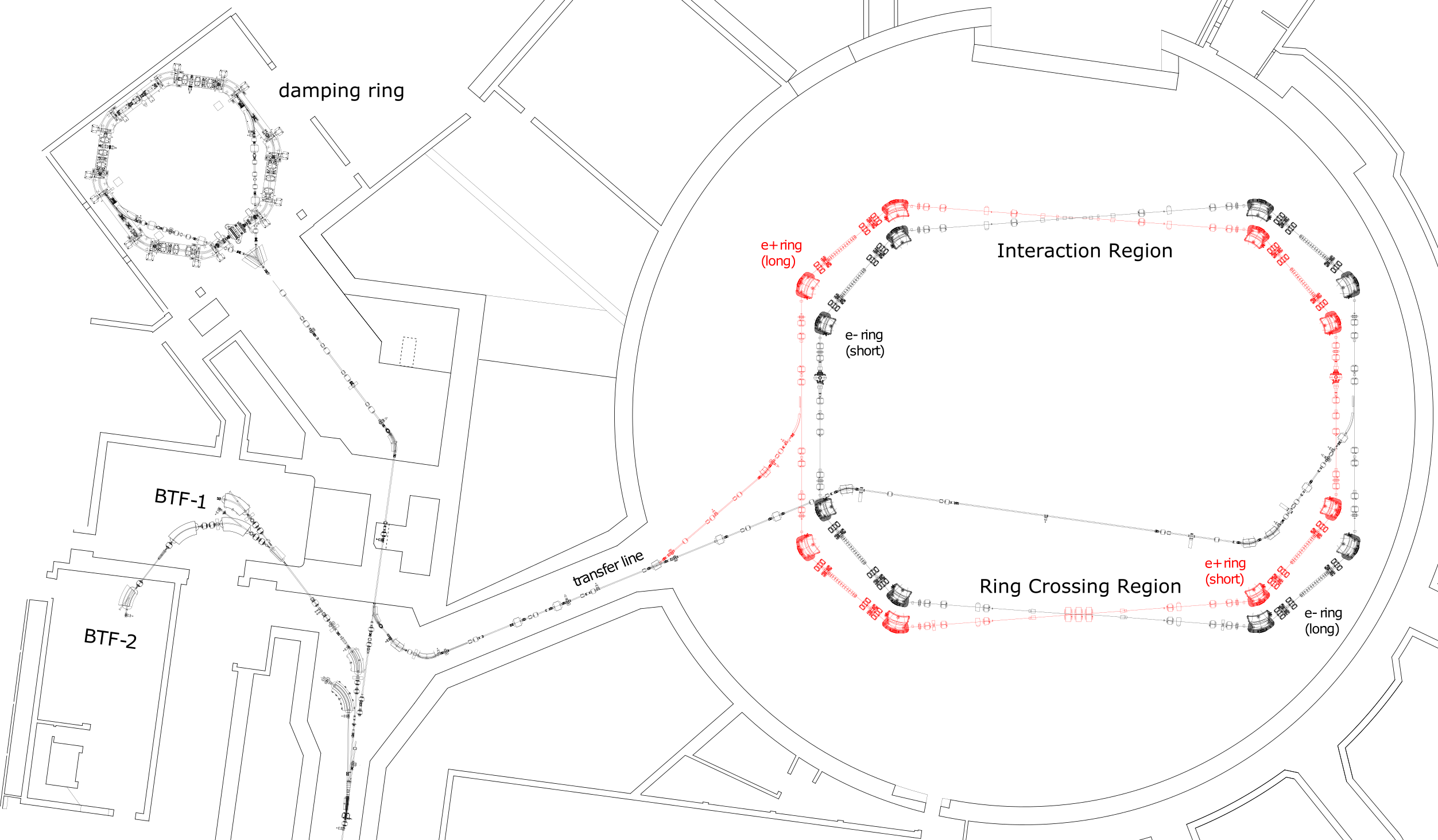}
  \caption{DA$\Phi$NE complex schematic layout: the beam from the LINAC (bottom left) can be driven either to the BTF-1 and BTF-2 lines (middle left) or to damping ring (top left) and from there injected  counter-clockwise into the electron main ring or clockwise into the positron one (in red).}\label{f:complex-layout}
\end{figure*}

The LINAC pulses can also be diverted to a separate transfer line where secondary beams can be produced on a dedicated target, serving a beam-test facility (BTF) with two experimental areas for high and medium-low intensity applications~\cite{Ghigo:2003gy,Valente:2016tom}.
A schematic layout from the end of the LINAC is shown in Fig.~ \ref{f:complex-layout}.\par
During the years 2018 to 2020, PADME has taken data using secondary and primary positrons in the BTF-1 beam-line; the main limitation to the sensitivity of the  experiment comes from the maximum tolerable pile-up in the detector, thus requiring to increase the positron beam  bunch length as much as possible.\par
\par
The DA$\Phi$NE LINAC was configured to produce pulses as long as 300~ns~\cite{Valente:2017mnr},
more than a factor ten longer than the design 10~ns required for injection into the collider, at the expense of a lower maximum accelerating field and higher energy spread; thus, beam pulses in the $0.43$ to $0.49$~GeV energy range were produced.\par
The maximum tolerable rate for the PADME detectors is of the order of $10^2$ positrons/ns, so that with 300~ns long pulses the maximum positron population cannot exceed $3\times10^4$ positrons/pulse. Estimating a year of operation as $10^7$~seconds at a repetition rate of 49~Hz, given by the LINAC system settings, it could be possible to achieve a maximum of $1.5\times10^{13}$ positrons on target. \par 
In order to further increase the reach of the experiment, it would be necessary to further extend the duration of the LINAC positron pulses.\par
As an alternative to LINAC modifications~\cite{Valente:2020fpo}, it has been proposed to use one of the rings of the DA$\Phi$NE complex as pulse stretcher~\cite{Valente:2017hjt}.\par
Two options were put forward: resonant one third of integer resonant extraction~\cite{Guiducci:2018koo} and \emph{ultra-slow extraction} using coherent effects in bent crystals. The latter is object of this paper, in which we report on the studies aiming at increasing the statistics of a PADME-like experiment by at least three orders of magnitude, i.e. extending the positron beam duration in the hundreds of $\mu$s range.\par Different options have been studied, 
%including using the crystal deflection in conjunction with the resonant technique,
also considering technical and practical aspects aiming at the design of a realistic and efficient implementation.

\section{The SHERPA extraction project}
The \emph{ultra-slow extraction} of a particle beam is performed by the turn--by--turn extraction of a small portion of the bunch population that is stored in an accelerator ring.\par
SHERPA (``Slow High-efficiency Extraction from Ring Positron Accelerator") is aimed at studying the possibility of ultra-slow positron extraction, aided by a bent crystal, from one of the three rings composing the DA$\Phi$NE complex: the Damping Ring~(DR), the electron main ring~(MRe) and the positron main ring~(MRp).\par

The crystal extraction is based on the possibility to deflect charged particles using the channeling effect in bent single crystals. High-energy charged particles impinging on the crystal with small angles relative to the lattice planes move oscillating between two neighboring planes, and consequently can be deflected by the crystal bend angle \cite{biryukov2013crystal}. 

This process and its application for slow extraction from ring accelerator have been studied and experimentally proved during the last decades. In particular, at the U-70 IHEP (``Institute of High Energy Physics") russian synchrotron \cite{afonin1998first} and at the CERN (``European Organization for Nuclear Research") SPS (``Super Proton Synchtron") \cite{Altuna1995, scandale2017possibility,Fraser2017experimental} crystal non resonant slow extraction has been successfully obtained for high-energy hadrons.

Applying the same technique in DA$\Phi$NE is challenging, but anyway possible. For sub-GeV leptons it is necessary to use ultra-thin crystals ($\sim$20 $\mu$m), instead of some mm, to reduce the interactions of crystal electrons on channeled particles pushing them out of the lattice potential well (``electron dechanneling"). 

We have obtained very promising preliminary results with few modifications at the current DA$\Phi$NE complex configuration. First, the inclusion of a crystal, and the structure to hold it and move it, into one of the vacuum chambers where the beam is circulating. Second, the angular deflection range needed, the crystal position, displacement at the extraction point, and initial considerations on the longitudinal position and geometry of the extraction septum.\par
Based on previous experimental results of particle channeling through bent crystals with electrons at about 1~GeV, reported in \cite{MAMI1,MAMI2}, the best performance of a silicon bent crystal of 30~$\mu$m thickness along the beam direction is about 1~mrad of deflection. The \emph{channeling efficiency}, defined as the percentage of particles deflected by channeling with respected to the total population, was measured to be at the level of $\approx20\%$ for electrons, and according to channeling theory efficiency for positrons is expected to be even higher.\par
The optical parameters of the rings were modified and tuned in simulations, using MAD-X~\cite{madx}, to allow the circulating beam to interact with the crystal and deflect particles for the slow extraction, while still allowing its injection and storage.\par
\begin{figure}
    \centering
    \includegraphics[width=0.35\textwidth]{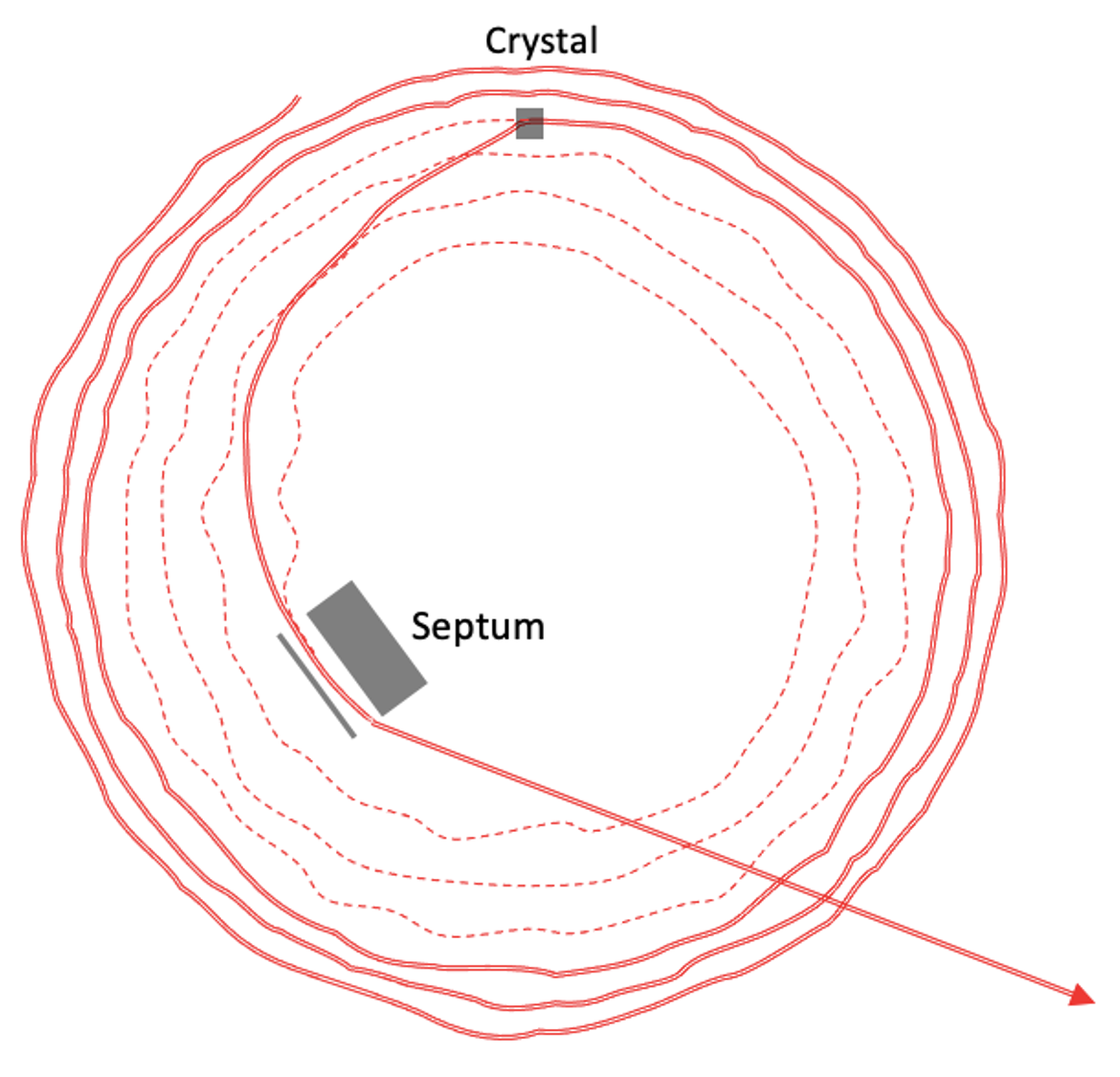}
    \caption{Non--local extraction of a particle using the deflection produced by channeling in a bent crystal. The particle oscillates in the storage ring until it interacts with the crystal, changing its trajectory towards an extraction section represented by a septum.}
    \label{f:nonlocal}
\end{figure}

%The ``local extraction'', in which the crystal and the extraction point are at the same location, is discarded because 1~mrad of deflection would need a few meters of free space to separate the extracted beam from the circulating by some millimeters (e.g. 10~mm~=~10~m~$\times$~1~mrad).\par
The most promising scheme was found in the so-called ``non--local'' crystal extraction: a deflection is imparted by a crystal at one point of the ring starting an oscillation and allowing particles to reach a septum with the adequate transverse displacement. Positrons can encounter the crystal multiple times and get further kicks or be lost (Fig.~\ref{f:nonlocal}).\par
In the non--local extraction, the transverse displacement $\Delta x_2$ at point 2 (the septum location) due to a kick $\Delta x'_1$ given at point 1 (the crystal location) can be calculated from the equation~\cite{widermann}
\begin{equation}
\Delta x_2 = \sqrt{\beta_1 \beta_2} \sin(2\pi\Delta\mu)\Delta x'_1, \label{eq:displacement}
\end{equation}
where $\beta_1$ and $\beta_2$ are the optics Twiss $\beta$ functions at the locations 1 and 2, $\Delta\mu$ is the phase advance between points 1 and 2 (in $2\pi$ units), and $\Delta x'_1$ is the deflection produced by the crystal. Equation~(\ref{eq:displacement}) can be seen as the displacement produced by the propagation of a kick given the linear optics transport element $R_{12}$ as explained by Wiedermann in~\cite{widermann}. In order to have maximum displacement $\Delta x_2$, the Twiss beta functions should be as large as possible and $\Delta \mu=\frac{1}{4}$.\par
In the following we describe the simulation and results obtained using the DA$\Phi$NE rings model for the two main options: using the smaller damping ring (A) and extracting from one of the two main rings (MRp in particular, B).\par

\subsection{Beam extraction from the damping ring}
The damping ring could provide an adequate extracted beam quality and could be considered as a good option because its structure is already used for beam extraction therefore requiring simpler modifications. In particular, the present extraction septum and the connected beam line can be used.\par
% There are  optimization due to the lower number of elements available for adaptability and tunability of the machine structure.\par
Particles bunches with a maximum population of $10^{10}$~electrons or $10^9$ positrons are accelerated by the LINAC to 510~MeV and injected into the 30~m long damping ring where the beam emittance is reduced by synchrotron radiation emission with a damping time of about 20~ms.\par% When beam damping is achieved, the beam is extracted and transported to one of the DAFNE Main Rings~(MR), one for electrons~(MRe) and another for positrons~(MRp), with a circumference of 100~m that are currently used for beam collisions at 1~GeV c.o.m. energy.\par

\par
\begin{figure}[htb]
\centering
\includegraphics[width=0.48\textwidth]{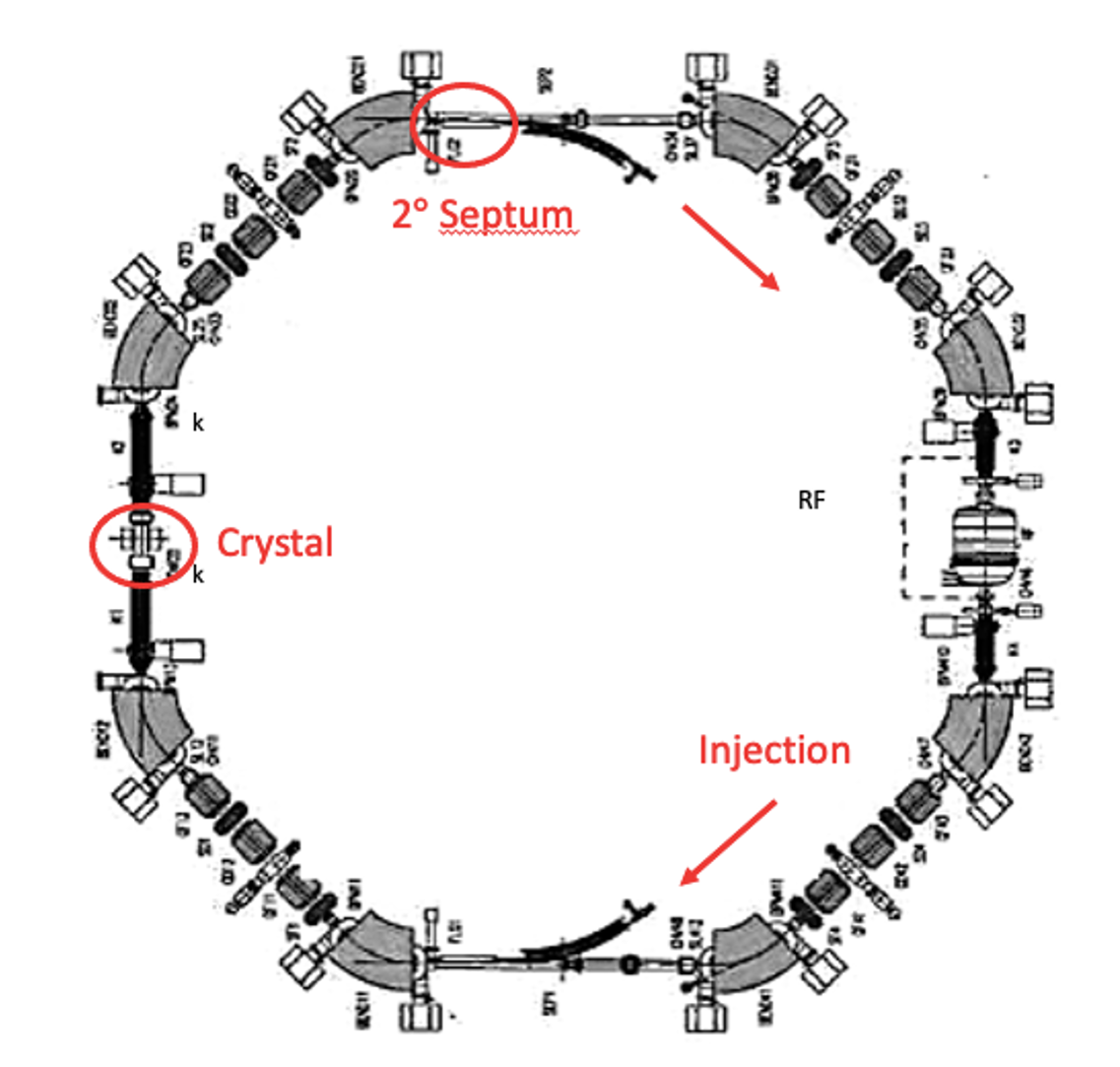}
\caption{Accumulator Ring, location of the bent crystal and extraction.}\label{f:dampingring}
\end{figure}

The proposed crystal location is shown in Fig.~\ref{f:dampingring}, together with 
 the extraction septum, which is the same used to extract the beam towards the DA$\Phi$NE main rings. \par
Particle tracking and theoretical estimations show that positrons with an energy of $-1.0$\%, with respect to the nominal one, will arrive at the crystal location with an horizontal displacement of $-7.5$~mm. Those particles could interact with the crystal and thus be kicked by about 1~mrad, producing a larger horizontal displacement of about $-11$~mm at the septum position. In order to  effectively achieve the slow extraction of those positrons it is necessary to adjust the initial position of the beam, the crystal position and the ring parameters.\par
Since positrons in the accumulator ring loose 1.0\% of energy by synchrotron radiation in 1000 turns, the extraction time is estimated to be 100~$\mu$s after injection at nominal energy, given that the RF (``Radio Frequency'') is kept off, thus allowing positrons to loose naturally energy and ``diffuse'' until they reach the crystal. Alternatively, one could modify and tune the extracted beam spill duration using the RF cavity for extending the time required to loose the same fraction of energy.\par
\begin{figure}[htb]
    \centering
    \includegraphics[width=0.48\textwidth]{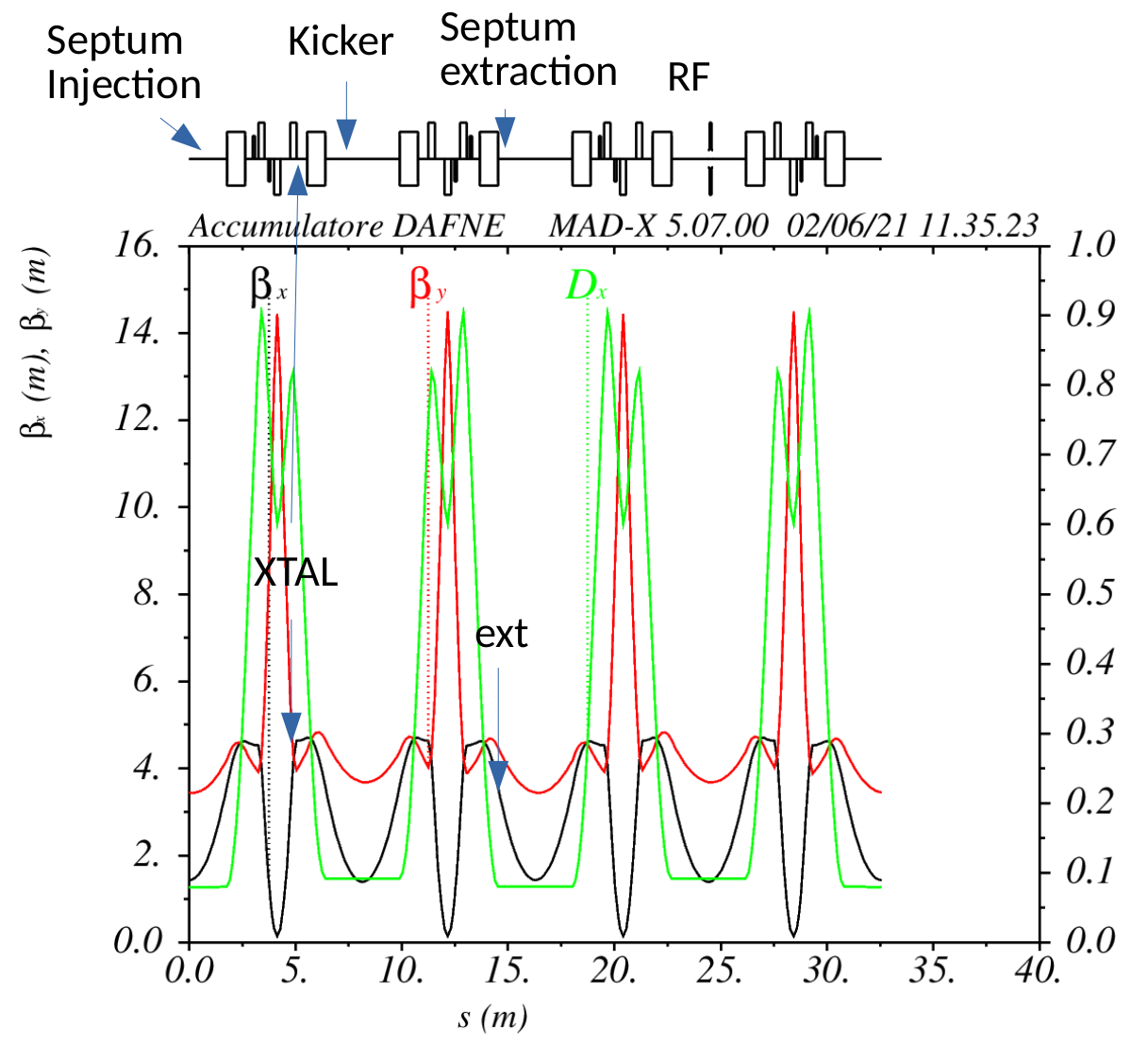}
    \caption{One possible configuration for the optics Twiss functions of the DA$\Phi$NE accumulator ring to allow crystal slow extraction.}
    \label{f:betaacc}
\end{figure}

In order to improve over the initial result, the optics Twiss functions have been modified to obtain a larger $\beta_x$ at the crystal and septum locations, while at the same time achieving a phase advance difference of 0.875 (in $2\pi$~units) between the two. Figure~\ref{f:betaacc} shows the Twiss functions, including the approximate location of the crystal and the extraction point.\par
We start the simulations considering a very small emittance, of 0.1~mm~mrad, in order to remove the effect of the beam size on the particle distribution. The energy spread of the beam is uniformly distributed between $\pm$1\%, which is the energy acceptance of the ring.\par
\begin{figure}[htb]
    \centering
    \includegraphics[width=0.48\textwidth]{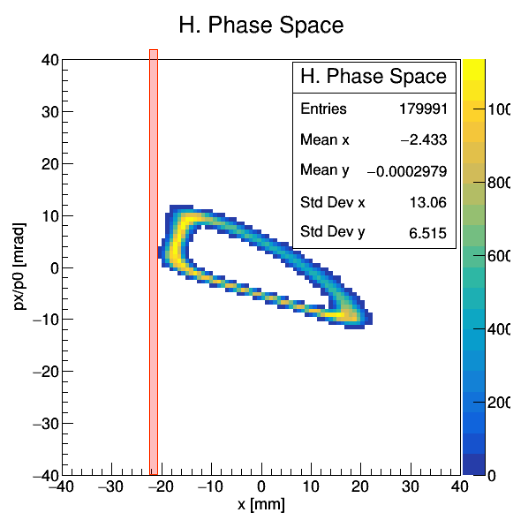}
    \caption{H Phase space plot  at the extraction septum of particles injected with a +10~mm offset from beam pipe center and tracked for 18 turns. The extraction septum is represented by a red region.}
    \label{f:phasespaceextract}
\end{figure}

%However, an estimation of the beam size of a particle bunch coming from the LINAC with an emittance below 5~mm~mrad and 1\% energy spread shows a maximum of 10~mm. Therefore, it is convenient to locate the crystal more than 1~cm away from the closed orbit.\par
Considering the septum located at $-20$~mm away from the beam pipe center, we bring the beam close to the septum by adjusting the kicker for off-axis injection. The injection oscillation brings the beam close to the septum, which is 2.5 mm thick, and the crystal provides the additional jump to pass the septum thickness. Figure~\ref{f:phasespaceextract} shows the distribution in the horizontal (H) phase space of particles at the extraction septum when they are injected 10~mm away from the beam pipe center and tracked for 18 turns.\par
 In the following we proceed to evaluate the effect of the bent crystal on the beam position at the extraction septum, using a beam deflection of $\pm$1~mrad, in agreement with the experimental results of a Silicon crystal $\sim 30~\mu$m thick \cite{MAMI1,MAMI2}.\par
\begin{figure}[h!]
    \centering
    \includegraphics[width=0.48\textwidth]{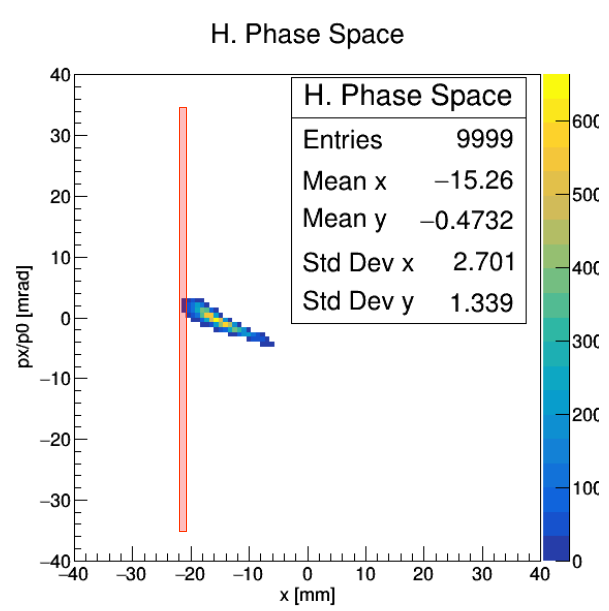}
    \includegraphics[width=0.48\textwidth]{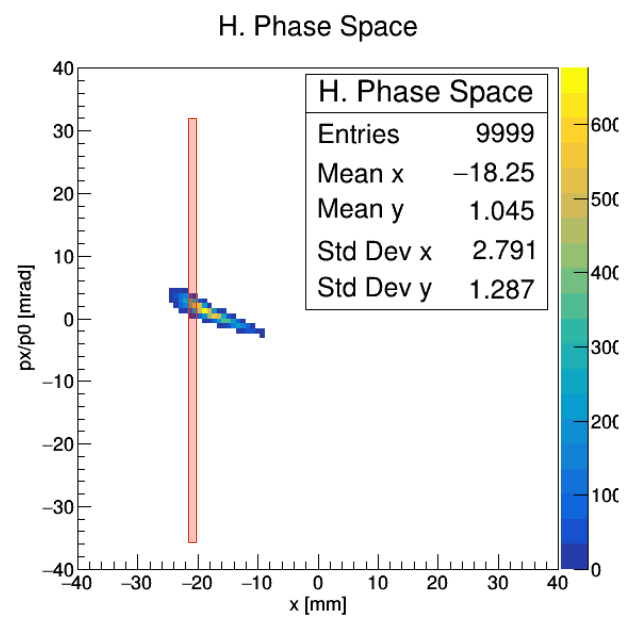}
    \caption{H Phase space plot at the extraction septum of particles injected with a +10~mm offset from beam pipe center and tracked for 1 turn. TOP: particles do not receive a kick by the crystal. Bottom: particles receive a 1~mrad kick by the crystal. Particles close to the septum are moved out of it, in the extraction channel.}
    \label{f:crystal}
\end{figure}

 
 As seen shown in Fig.~\ref{f:crystal} the effect of the crystal at the first turn is to displace the particles by about 4.0 mm at the extraction septum. The position of the crystal is chosen in order to interact with the particles which are very close to the extraction septum in order to extract them with the 1~mrad kick.  The crystal is placed at -3 cm from the beam center in the horizontal plane.\par
\begin{figure}[h!]
    \centering
    \includegraphics[width=0.48\textwidth]{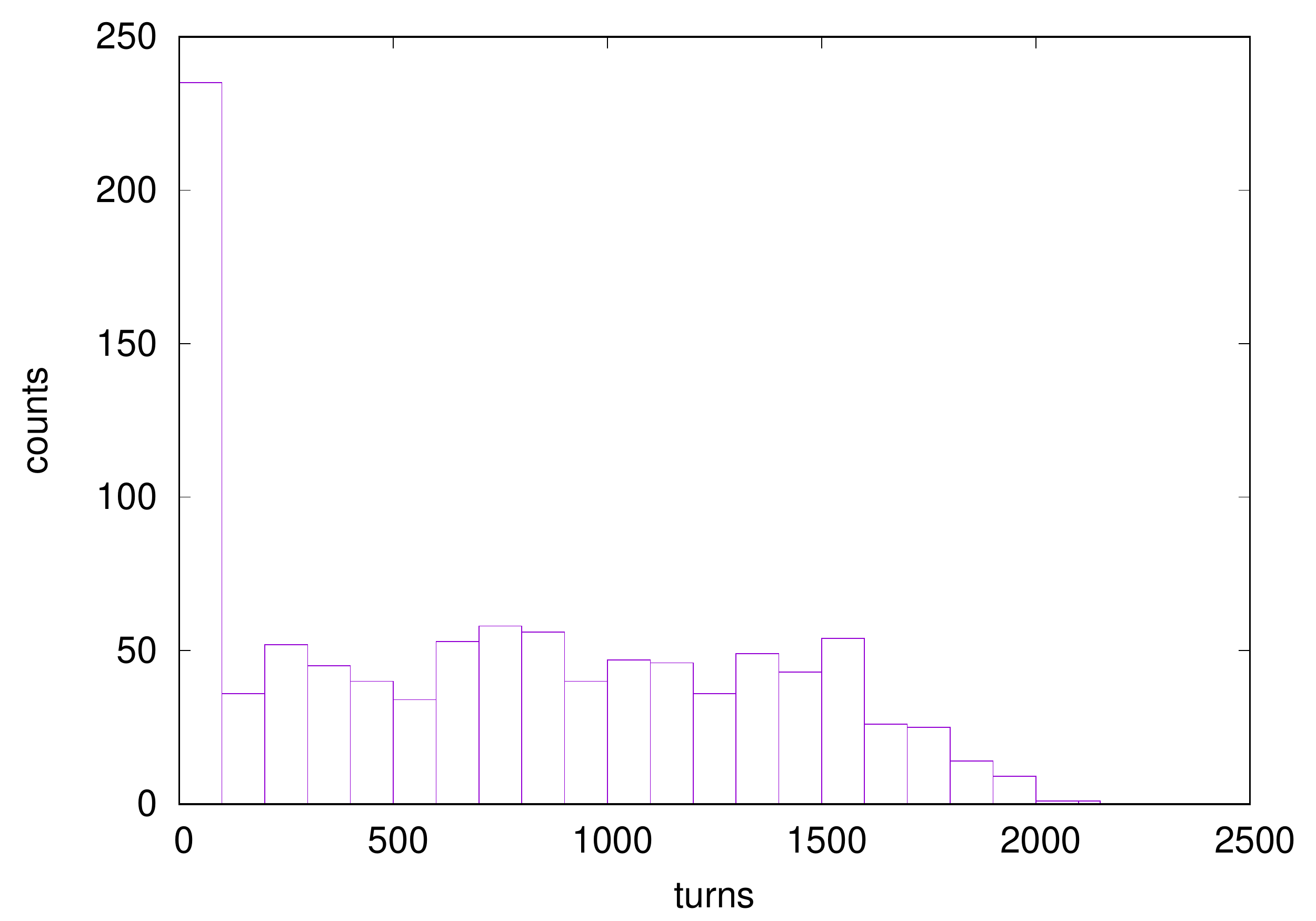}
    \caption{Rate of particles achieving  $-20$~mm of displacement, or more, at the extraction point. The initial tracked population is $10^3$ particles and each turn corresponds to 0.1~$\mu$s. Each bin corresponds to 100 turns.}
    \label{f:rate}
\end{figure}

The rate of particles passing through the crystal at $-30$~mm and achieving a displacement of $-20$~mm or more at the extraction septum is plotted in Figure~\ref{f:rate}. There is a peak in the rate at the first turn, which can be reduced by further optimization of the initial conditions. After the first peak the rate of particles is rather stable up to about 1500 turns.\par
\begin{figure}[h!]
    \centering
    \includegraphics[width=0.48\textwidth]{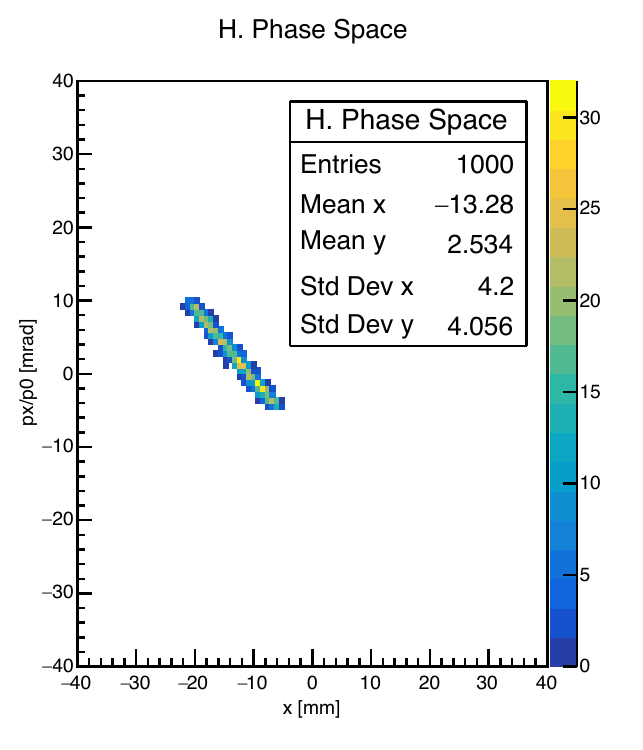}
    \includegraphics[width=0.48\textwidth]{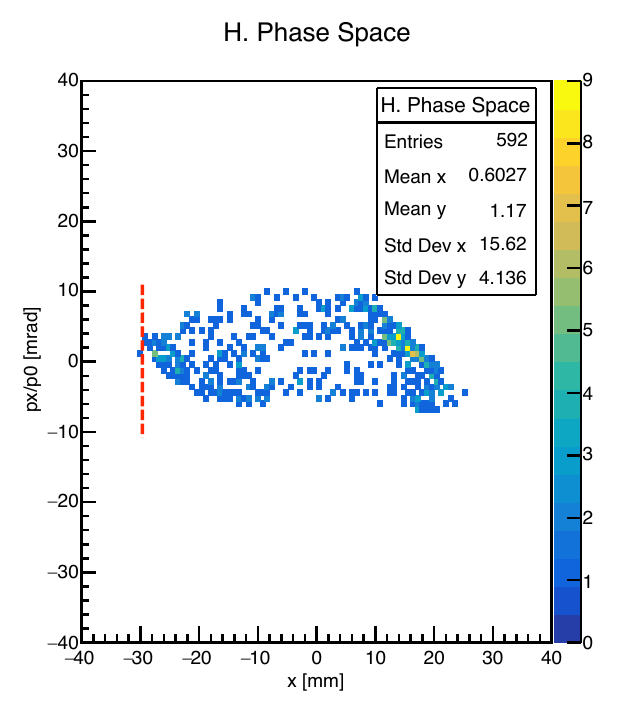}
    \caption{H Phase space of particles achieving $-2$~cm of displacement, or more, at the extraction point plotted at the crystal. TOP: turn 1. BOTTOM: turn 500. Particles migrate to negative horizontal offsets due to dispersion and energy loss up to go through the crystal (represented by the dashed red line in the plot).}
    \label{f:migration}
\end{figure}

 In Fig~\ref{f:migration} the phase space of the same particles seen at the crystal location is shown at the first turn and at turn 500. The particles are tracked until they reach the position of $-20$~mm or more at the extraction septum. After $\sim$ 500 turns the first particles pass through the bent crystal receiving the 1 mrad kick. \par

\subsection{Simulation studies of the DA$\Phi$NE collider positron beam extraction with a bent crystal}
The DA$\Phi$NE collider is composed by two rings and the structure of both main rings is similar. Any of the two has great flexibility due to the numerous devices (quadrupoles, kickers, beam monitors, sextupoles, etc.) to manage and control the beam parameters for an optimised beam quality in terms of intensity, emittance and rate of extraction. However, this option requires greater efforts in terms of human resources, design, installation, commissioning and operation. Differently from the accumulator ring options, the minimum set of modifications indeed includes the installation of the extraction septum and a suitable transport line.\par
Starting from the DA$\Phi$NE collider configuration used for the SIDDHARTA2~(2019)~\cite{siddharta} experiment, we have modified the optics model, finding different promising options for the crystal ``non-local'' extraction.\par
\begin{figure}[htb]
    \centering
    \includegraphics[width=0.48\textwidth]{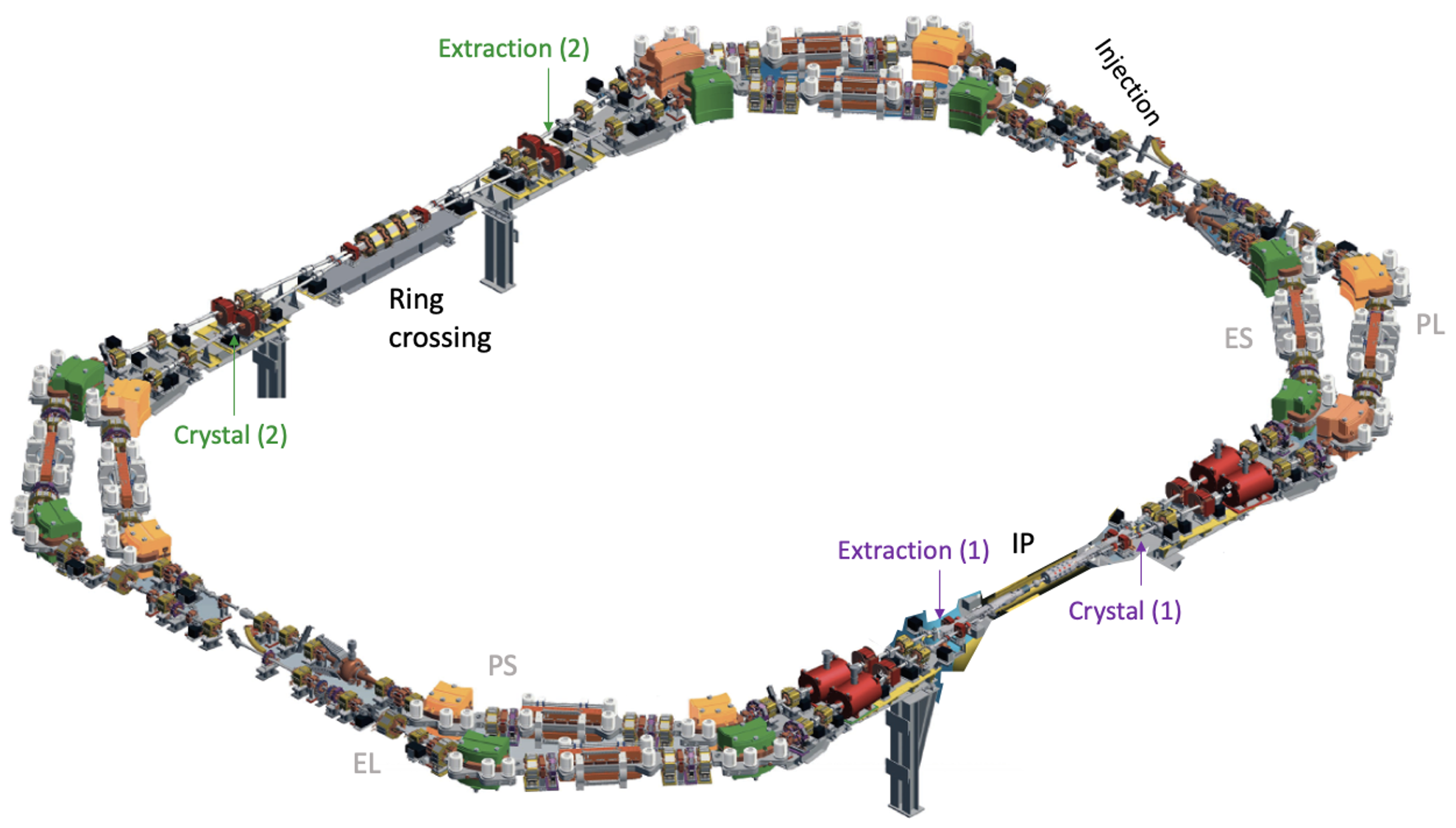}
    \caption{Non local extraction scheme locations using the DA$\Phi$NE positron main ring; particles are injected in the long external arc (PL) and circulate clockwise: in option (1) the crystal and the septum are placed upstream and downstream the main experiment interaction point, in case (2) they are placed around the ring crossing region on the opposite side.}
    \label{f:dafnemrpip2}
\end{figure}

The best configuration is the one with the crystal positioned just before the rings crossing point~(IP2), shown in Fig.~\ref{f:dafnemrpip2}, and the extraction septum placed just downstream the IP2. The extraction would be thus performed only a few meters downstream the crystal, in the same straight section of the MRp ring.\par
\begin{figure}[h!]
    \centering
    \includegraphics[width=0.48\textwidth]{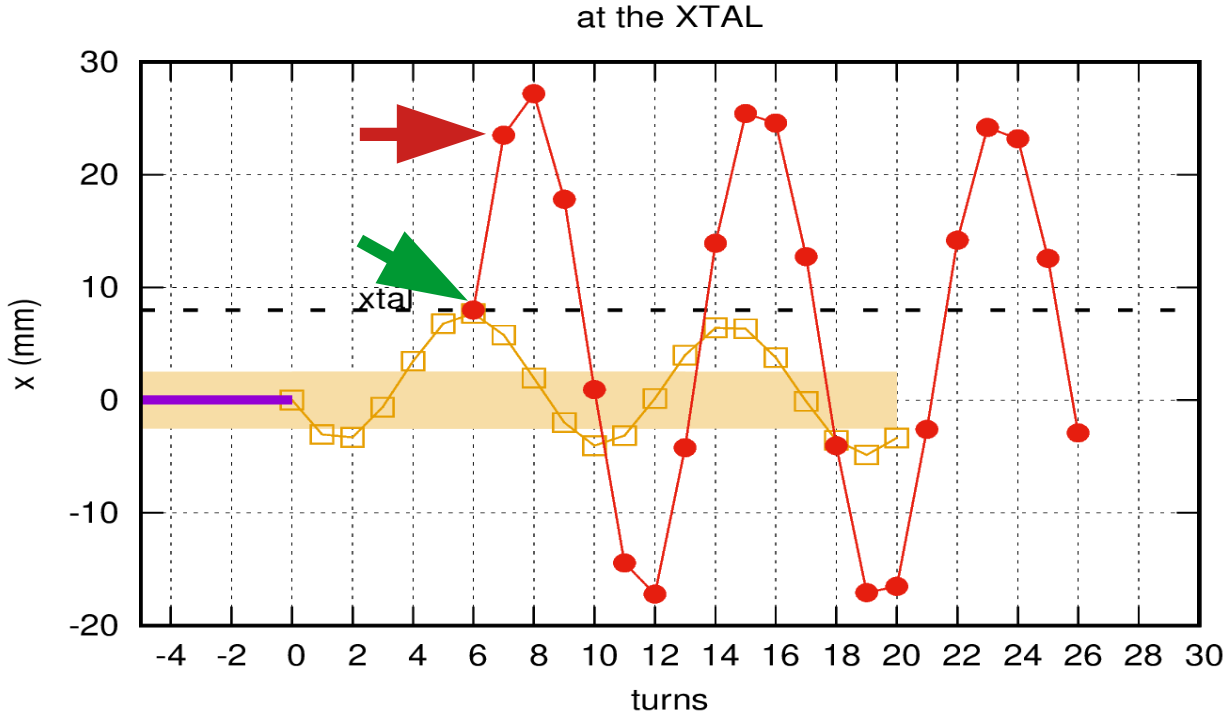}
    \includegraphics[width=0.48\textwidth]{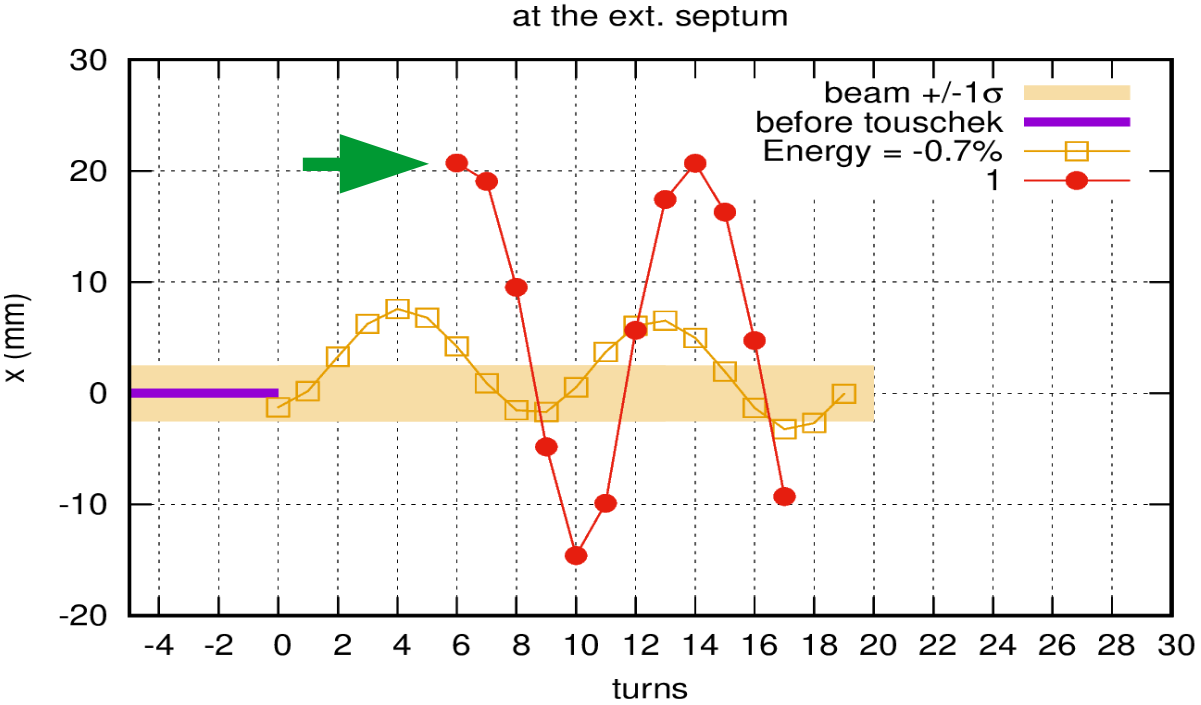}
    \caption{Turn by turn single particle tracking results seen at the Crystal location~(TOP) and the Extraction Septum~(BOTTOM). The particle trajectory is initially set to zero position and angle and at turn zero it loses 0.7\% of its energy. At turn 6 it interacts with the crystal and it reaches 20~mm of displacement at the septum. If not extracted, the particle continues to oscillate with a large amplitude.}
    \label{f:turnbyturn}
\end{figure}

In this configuration, a positron with an energy offset of the order of -0.7\% will encounter the crystal, positioned at 8~mm from the circulating beam axis, at the 6$^{th}$ turn in
the machine, and will be extracted in the same turn, as shown in Fig.~\ref{f:turnbyturn}.\par
The transverse displacement obtained at the extraction point is about 20~mm due to the large horizontal beta~$\beta_x$ of about 17~m at both locations, enough to reach the typical offset separating the beam and the extraction septum. 
\par
It has to be underlined that the energy loss considered in the simulation is realistic as several phenomena could change a particle energy, e.g. synchrotron radiation, inelastic beam-gas scattering or the intra--beam particle interaction (the so-called Touschek effect). In any of the previous cases particles losing the right amount of energy will be deflected by the bent crystal and slowly extracted with the help of a standard septum.\par
One could consider the possibility of forcing the migration of the bunch population to a negative energy offset by turning off the DA$\Phi$NE RF device, or even slow down this migration process by using the cavity to recover only part of the energy radiated by the bunch and at the same time to keep the beam stable along the ring.\par
In any case, this preliminary study already shows that it is possible to slow extract positrons from the actual DA$\Phi$NE ring using a bending crystal with a realistic deflection, i.e. already achieved with a lepton beam of comparable momentum.\par

\section{Future work and conclusions}
The crystal extraction solutions proposed here have to be optimised in terms of injection and optics parameter, RF tuning, crystal position and septum features to obtain the best result in terms of extracted beam quality.\par
In particular the extracted beam main parameters to be optimised are: spill length, intensity, emittance and energy spread.\par
For the main ring configuration, a resonant solution has been also investigated \cite{Guiducci:2018koo} and it could be implemented together with the crystal solution for a hybrid ``resonant-non resonant'' extraction approach. This option is still under study.\par
Also for the damping ring, a resonant solution and a hybrid ”resonant-non resonant” crystal extraction is under investigation.\par
Monte Carlo studies and analytical simulation studies, especially on the particles behaviour in the crystal, are ongoing with the aim to adapt the existing codes optimised for high-energy hadrons also to sub-GeV leptons. The data that will be collected characterising the SHERPA crystals, will be also used for simulation benchmark.

It is important to underline that similar results are also expected for electron beams, albeit with an expected lower crystal deflection efficiency.

If SHERPA will succeed in its final objective of achieving the crystal-assisted extraction, the very first sub-GeV primary positron slowly-extracted beam will be delivered. This would open the possibility of managing low-energy positron and electron beams in a small storage ring. The possibility to extract high-quality spills of O(ms) will guarantee new scenarios for several fixed-target positron experiments: reducing the pile-up, the background, the energy spread, their sensitivity will greatly increase the physics reach of PADME-like experiments. Moreover, the study of positron beam steering using bent crystals will provide a know how that can be applied, in the next future, for several accelerating machine aspects, like beam collimation, extraction and splitting, contributing to a general improvement in the particle accelerator field. For the $DA\Phi$NE complex, this project could be an important chance to upgrade its performance, widening its use-cases in different research line, also for fundamental physics.

\section{Acknowledgments}
This work has been financially supported by the Istituto Nazionale di Fisica Nucleare~(INFN), Italy, Commissione Scientifica Nazionale~5, Ricerca Tecnologica -- Bando n.~21188/2019.\par

\newpage
%\bibliography{biblio}
\bibliographystyle{unsrt}

\end{document}